# Electron irradiation and thermal chemistry studies of interstellar and planetary ice analogues at the ICA astrochemistry facility

Duncan V. Mifsud[1,2], Zoltán Juhász[2*], Péter Herczku[2], Sándor T. S. Kovács[2], Sergio Ioppolo[3], Zuzana Kaňuchová[4,5], Máté Czentye[1], Perry A. Hailey[1], Alejandra Traspas Muiña[3], Nigel J. Mason[1], Robert W. McCullough[6], Béla Paripás[7], Béla Sulik[2]

1       Centre for Astrophysics and Planetary Science, School of Physical Sciences, University of Kent, Canterbury CT2 7NH, United Kingdom

2       Institute for Nuclear Research (Atomki), Debrecen H-4026, PO Box 51, Hungary

3       School of Electronic Engineering and Computer Science, Queen Mary University of London, London E1 4NS, United Kingdom

4       Astronomical Institute of the Slovak Academy of Sciences, Tatranska Lomnicá SK-059 60, Slovakia

5       INAF Osservatorio Astronomico di Roma, Monte Porzio Catone RM-00078, Italy

6       Department of Physics and Astronomy, School of Mathematics and Physics, Queen's University Belfast, Belfast BT7 1NN, United Kingdom

7       Department of Physics, Faculty of Mechanical Engineering and Informatics, University of Miskolc, Miskolc H-3515, Hungary

\*      Corresponding author:        zjuhasz@atomki.mta.hu

## Author ORCID Identification Numbers:

| | |
|---|---|
| DVM | 0000-0002-0379-354X |
| ZJ | 0000-0003-3612-0437 |
| PH | 0000-0002-1046-1375 |
| STSK | 0000-0001-5332-3901 |
| SI | 0000-0002-2271-1781 |
| ZK | 0000-0001-8845-6202 |
| PAH | 0000-0002-8121-9674 |
| ATM | 0000-0002-4304-2628 |
| NJM | 0000-0002-4468-8324 |
| BP | 0000-0003-1453-1606 |
| BS | 0000-0001-8088-5766 |

**Abstract**

The modelling of molecular excitation and dissociation processes relevant to astrochemistry requires the validation of theories by comparison with data generated from laboratory experimentation. The newly commissioned Ice Chamber for Astrophysics-Astrochemistry (ICA) allows for the study of astrophysical ice analogues and their evolution when subjected to energetic processing, thus simulating the processes and alterations interstellar icy grain mantles and icy outer Solar System bodies undergo. ICA is an ultra-high vacuum compatible chamber containing a series of IR-transparent substrates upon which the ice analogues may be deposited at temperatures of down to 20 K. Processing of the ices may be performed in one of three ways: (i) ion impacts with projectiles delivered by a 2 MV Tandetron-type accelerator, (ii) electron irradiation from a gun fitted directly to the chamber, and (iii) thermal processing across a temperature range of 20-300 K. The physico-chemical evolution of the ices is studied *in situ* using FTIR absorbance spectroscopy and quadrupole mass spectrometry. In this paper, we present an overview of the ICA facility with a focus on characterising the electron beams used for electron impact studies, as well as reporting the preliminary results obtained during electron irradiation and thermal processing of selected ices.

## 1 Introduction

The modelling of intense excitation processes in low-temperature ices has found applications in a wide variety of fields [1], but particularly in molecular astrophysics where such processes may lead to novel chemistry within the ice structure [2-5]. Interstellar and planetary ices may be modelled as dense gases experiencing weak intermolecular forces of attraction and restricted degrees of freedom. Although astrochemical modelling on its own may greatly contribute to our knowledge of extra-terrestrial chemistry, it is often necessary to perform comparative laboratory experiments so as to serve as a benchmark against which multi-scale models and theories may be tested and validated.

Accordingly, the establishment of experimental facilities where solid-phase astrochemistry may be investigated is an important aspect of the development of astrochemistry research. The recently commissioned Ice Chamber for Astrophysics-Astrochemistry (ICA), hosted by the Institute for Nuclear Research (Atomki) in Debrecen, is one such facility which is able to simulate various astrophysical environments (e.g. the interstellar medium, icy planetary and lunar surfaces, etc.) and quantitatively analyse the physico-chemical changes occurring in deposited astrophysical ice analogues as a result of energetic processing.

Within the context of interstellar and Solar System ice astrochemistry, one of the more notable forms of energetic processing is electron irradiation. The study of electron-induced chemistry in astrophysical environments is important as it is thought that such chemistry is a major route to the synthesis of molecules [6,7]. Non-thermal low-energy (<20 eV) electrons are produced as a result of the interaction between ionising radiation and matter and are believed to drive most of the radiolytic chemistry in astrophysical ices via a combination of impact ionisations, electronic excitations, and dissociative electron attachments [8]. Higher energy electrons are a component of galactic cosmic rays, planetary magnetospheric plasmas, and the solar wind [8-12] and their impact into ices may also engender radiochemical reactions via the ionisation of the atomic and molecular constituents of the ice.

Another significant form of processing to which astrophysical ices are subjected is thermal processing. Thermally-induced chemical reactions may occur in all astrophysical settings where temperatures are high enough to overcome the relevant activation energy barriers [13]. This is especially important in the contexts of comets and icy outer Solar System moons, where such chemistry is not only known to be prevalent, but also leads to the formation of complex molecules of astrobiological relevance. For instance, the thermal processing of ices containing ammonia ($NH_3$), methanimine ($CH_2NH_2$), and hydrogen cyanide (HCN) has been shown to yield aminoacetonitrile ($NH_2CH_2CN$); an important precursor to amino acids [14].

The importance of characterising the electron irradiation and thermal processing leading to interstellar and outer Solar System ice chemistry is therefore apparent. In this paper, we present a brief introduction to ICA so as to highlight some of its experimental capabilities with a focus on characterising the electron beam profiles used for electron irradiation studies. Additionally, we also present the results of the electron irradiation of amorphous methanol ($CH_3OH$) ice at 20 K and the thermal processing of an ice mixture composed of water ($H_2O$) and sulphur dioxide ($SO_2$) in order to further demonstrate the ability of the set-up to provide data which may be useful to the astrophysical and astrochemical modelling communities.

## 2 Description of the ICA facility

A complete technical description of ICA is provided in another publication [15], and thus in the present paper we limit ourselves to a brief overview. ICA is composed of an ultra-high vacuum compatible chamber of inner diameter 160 mm, within the centre of which is a heat-shielded copper sample holder (Fig. 1). This sample holder currently may hold up to four substrates onto which ice analogue replicates may be deposited under identical conditions. Thus, ICA has been purposefully designed to facilitate the performance of systematic ice processing studies where a number of experimental parameters (e.g. ice thickness, morphology, temperature, processing type, etc.) may be controlled and varied with ease.

Pressure within the chamber may be reduced to a few $10^{-7}$ mbar with the combined use of a dry rough vacuum pump and a turbomolecular pump. Even lower pressures of a few $10^{-8}$ mbar are attained upon cooling of the sample holder, which is performed using a closed-cycle helium cryostat able to offer a working temperature range of 20-300 K. Accurate temperature measurements are made using two silicon diodes connected to a Lake Shore temperature controller and a proportional integral-differential controller. The positioning of one diode at the top of the sample holder and the other at the bottom allows for the identification of any potential temperature gradients across the holder which may introduce uncertainties during experimentation.

The deposition of astrophysical ice analogues onto the substrates is conducted by introducing gases into the chamber via a fine needle valve. The presence of a distributor (scattering) plate in front of the inlet tube allows for a reduction in chamber pressure heterogeneity during this background deposition, thus ensuring that the ices produced are of roughly the same thickness on all deposition substrates. Both uni- and multi-component ices may be prepared by making use of a system of valves to introduce the gases into a mixing container, the partial pressures of which are monitored by a mass independent capacitive manometer gauge.

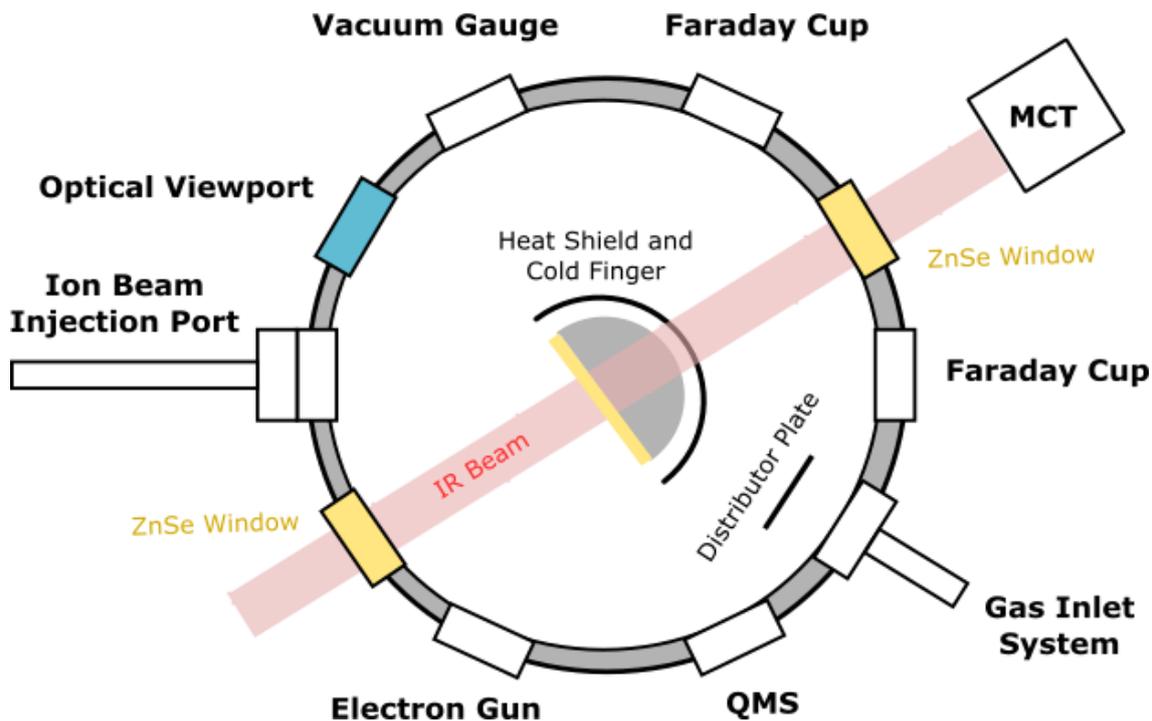

**Fig. 1** Top-view schematic diagram of the ICA chamber. Although the sample holder and heat shield are rotatable, both ion beam and electron beam irradiations are typically performed as depicted, with the IR beam pathway orthogonal to the sample surface and charged projectiles impacting at angles of 36°.

The chamber is equipped with ten DN-40 CF ports on its side walls separated from one another by angles of 36° which are used for external connections (Fig. 1). One of these ports hosts a Kimball ELG-2A electron gun for electron irradiation studies. The emitted electron energy range of this gun is 5-2000 eV, and beam current stability and intensity can be monitored using a Faraday cup mounted on the port directly opposite to the electron gun. For this monitoring to take place, the sample holder includes a 9.6 mm diameter collimator in place of one of the deposition substrates. The characterisation of the emitted electron beam is discussed in Sect. 3, while results obtained from the radiolysis of $CH_3OH$ ice are presented in Sect. 4.

The port located at 72° from the electron gun in a clockwise sense serves as the entrance for projectile ion beams supplied by a 2 MV Tandetron-type accelerator; the stability and intensity of these beams may be monitored using another Faraday cup mounted on the port directly opposite to the ion beam entry port (Fig. 1). The accelerator facility, as well as preliminary experimental results obtained by ion radiolysis of astrophysical ice analogues, are the focus of separate publications [15,16]. Other ports on the chamber are used to house the gas inlet dosing line, a vacuum gauge, and an optical viewport for direct observation of the sample holder and substrates.

Another two ports on the chamber are used as the entrance and exit points for the IR beam used to monitor the physico-chemical evolution of the interstellar and Solar System ice analogues undergoing energetic processing (Fig. 1). The ice samples are analysed by FTIR absorbance spectroscopy with IR analysis nominally set up in transmission mode using IR-transparent deposition substrates (typically zinc selenide). These substrates are coated with a fine gold mesh to prevent charging of the surface during high-current (>100 nA) irradiation [15]. Prior to ice deposition, background spectra of the bare substrates are obtained at the appropriate temperature and pressure and subtracted from the spectra collected during processing. During irradiation, species may be sputtered or desorbed from the ice and these molecules may be analysed by means of a quadrupole mass spectrometer located on another port.

## 3 Electron beam analysis

### 3.1 Electron beam profiling and flux determination

In order to qualitatively assess any chemical changes brought about by electron irradiation of deposited ice layers, it is necessary to have knowledge of both the current density distribution at the surface of the ice as well as the electron flux. Optimum conditions for studying the physico-chemical evolution of the ice are attained when the current density is constant over the entirety of the ice surface which is monitored by the IR spectroscopic beam. However, a nearly constant current density within a circular area which is smaller than that monitored by the IR beam may also be used, provided that appropriate measurement corrections are made (an example of which is discussed in more detail in Sect. 3.2).

To characterise the electron beam profile, a 9.6 mm diameter collimator is mounted onto the sample holder in place of one of the deposition substrates. The emitted electron beam is passed through this collimator and into the Faraday cup opposite the electron gun (Fig. 1), which is used to measure the current as a function of the vertical position of the collimator $Y$ as it is displaced from its nominal position in fine steps. The zero-value for the $x$, $y$, and $Y$ coordinates at the surface of the ice is defined by the passage of the axis of the electron beam through the plane of the surface. The measured current values $I(Y)$ may then be compared to model calculations:

$$I_{\text{model}}(Y) = \iint\limits_{x^2+y^2 \leq R_c^2} dx\, dy\, i(x, y-Y)$$

(Eq. 1)

where $R_c$ is the radius of the collimator and $i(x, y)$ is the hypothetical current density in the plane of the ice sample surface. A unique solution for the model function current density $i$ does not exist when compared with measured values and so, in the strictest sense, this cannot be considered a fit. However, given reasonable, few-parameter $i$ functions, a reasonable estimation for the shape and uniformity of the beam may be obtained. An example of this for a 2 keV electron beam (as was used in Sect. 4) is given in Fig. 2 which shows that our measured data matches the current density function for a modelled cylindrical and homogeneous beam of diameter 8.4 mm quite well.

The characterisation of the electron beam profile allows for the determination of the beam spot area $A$ incident on the sample ice during irradiation. This value, together with the measured current $I_{\text{max}}$ and the fundamental electric charge ($e = 1.602 \times 10^{-19}$ C) may be used to calculate the electron flux $\Phi$ (electrons cm$^{-2}$ s$^{-1}$). In order to ensure that a constant flux is used throughout an experiment, the beam current is measured on the Faraday cup opposite to the electron gun prior to each irradiation and compared to the current emitted by the filament, which is displayed on the digital power supply system. This displayed emitted current was continuously monitored during irradiation and was observed to be stable. Our test runs showed that the ratio between the measured and displayed currents did not change by more than 1% over five hours. By integrating the flux over the time of irradiation $t$, the electron fluence $\varphi$ (electrons cm$^{-2}$) may be determined:

$$\Phi = \frac{I_{\max}}{A \times e}$$

(Eq. 2a)

$$\varphi = \int_{t=0}^{t} \Phi \, dt$$

(Eq. 2b)

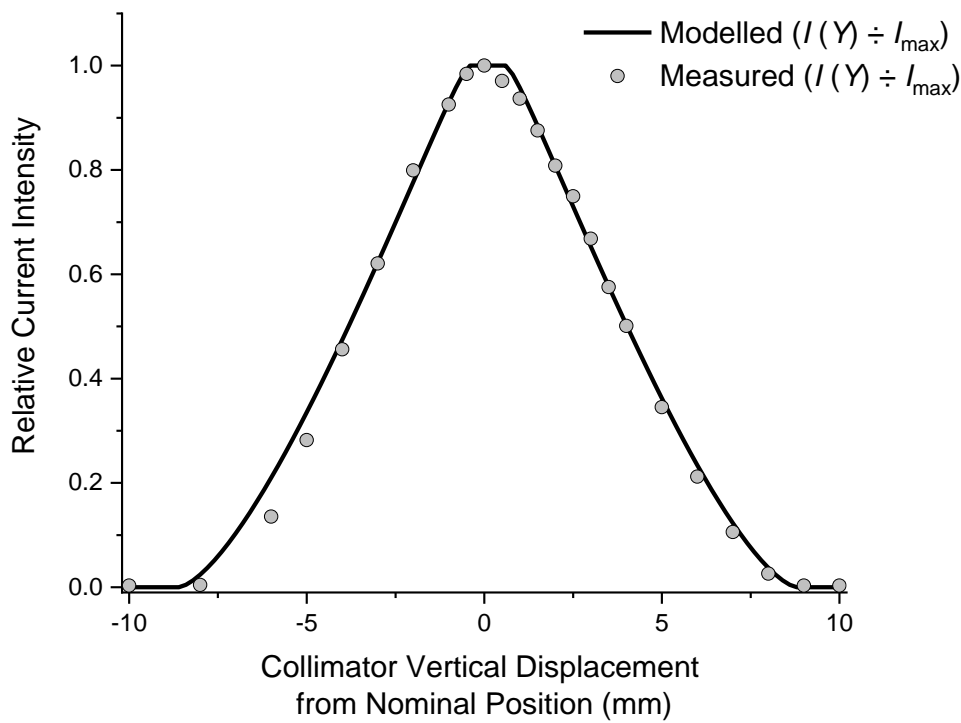

**Fig. 2** Plots of measured electron current $I(Y)$ relative to maximum beam current $I_{\max}$ as a function of the displacement of the 9.6 mm diameter collimator from its nominal position along the vertical axis for a 2 keV electron beam. In these plots, points represent measured values, while the plotted line is defined by Eq. 1 for a simple, cylindrical, homogeneous electron beam with a diameter of 8.4 mm at the sample ice surface.

Fig. 2 represents a rather special electron beam focusing condition which cannot be performed for all cases. Sharp focusing with a beam diameter <3 mm, however, can be set at all beam energies and can be checked using Eq. 1. In general, we ensure a uniform electron beam density by sweeping this focused electron beam in the $x$ and $y$ directions using sawtooth-shaped voltages applied to the deflection electrodes of the electron gun. In such cases, the uniformity of the beam current density can also be determined from the measured profile and the sample geometry by using Eq. 1 with a straightforward algorithm. Making use of this sweeping mode is the preferred option if beam homogeneity is important and the beam current density is the relevant physical quantity. If an accurate value of the total flux is more important, a beam spot size <9.6 mm (as depicted in Fig. 2) is the optimum solution.

## 3.2 Corrections to measured molecular column densities

In circumstances where the surface area of the ice irradiated by electrons spans the entirety of the area scanned by the IR monitoring beam, the abundances of both reactant and product molecules may be determined spectroscopically by measuring the peak areas of their characteristic absorbance bands and calculating the column density $N$ (molecules cm$^{-2}$):

$$N = \frac{\ln(10) \times P}{A_\nu}$$

(Eq. 3)

where $P$ is the peak area of the characteristic band for the given molecular species (cm$^{-1}$) and $A_\nu$ is the integrated band strength value for that particular band (cm molecule$^{-1}$) [17]. These column densities may be normalised as a fraction of the initial (pre-irradiative) column density of the reactant species $N_0$:

$$n = \frac{N}{N_0}$$

(Eq. 4)

However, it may be the case that the overall surface area irradiated by the electron beam is smaller than that monitored spectroscopically. Furthermore, the range (penetration depth) of the electrons may be smaller than the actual thickness of the target ice. In such scenarios, there exists an 'active volume' within the ice in which electron-induced chemistry is taking place. As such, it is preferable to correct for the 'inactive volume' when making measurements of the relative abundances of reactant and product molecules present. Under prolonged irradiation, it is possible to assume that the overwhelming majority of reactant molecules within the active volume are destroyed. Therefore, a working approximation of the ratio of this active volume to the volume monitored by the IR beam may be gleaned from the calculated final normalised column density $n_{\text{fin}}$ at the end of irradiation. This allows for the following corrected column density relations to be proposed for reactant $x$ and product $y_i$ molecules:

$$x = \frac{c}{c_0} = \frac{n - n_{\text{fin}}}{1 - n_{\text{fin}}}$$

(Eq. 5a)

$$y_i = \frac{c_i}{c_0} = \frac{n_i}{1 - n_{\text{fin}}}$$

(Eq. 5b)

where $n$ and $n_i$ refer to the normalised column densities of the respective reactant and product species, and $c$ and $c_i$ refer to the molecular concentrations of the reactant and product species, while $c_0$ refers to the initial molecular concentration of the reactant prior to irradiation. The extension of these corrections to mixed ices is relatively straightforward. The advantage of Eqs. 5a and 5b is that they provide a model independent, one parameter correction which is reasonable under prolonged irradiation. The more complicated general formulae (based on the Beer-Lambert Law) are not given here.

## 4 High-energy electron irradiation of amorphous $CH_3OH$ ice

In this Sect., we provide some preliminary results from the 2 keV electron irradiation of amorphous $CH_3OH$ ice at 20 K in order to showcase some of the experimental capabilities of ICA. Our choice of $CH_3OH$ as a target ice was motivated by the fact that this species is one of the more common extra-terrestrial molecules, having already been identified in several astrophysical environments including low- and high-mass protostars, comets, and centaur planetoids [6,8]. Furthermore, several studies have shown that the radiolytic processing of $CH_3OH$ may yield complex organic molecules, many of which are relevant to astrobiology (for a detailed review on the radiochemical processing of $CH_3OH$, see [8]).

The experimental protocol followed for this irradiation was as follows: an aliquot of $CH_3OH$ was de-gassed in a glass vial via several freeze-thaw cycles using liquid nitrogen. Vapours from this aliquot were subsequently introduced into the chamber at 20 K to deposit a 1 μm ice on zinc selenide substrates. The thickness $d$ (μm) of the ice was determined using the measured column density $N$ at the 1027 cm$^{-1}$ absorbance band and the known molecular mass $m$ (u) and density $\rho$ (g cm$^{-3}$) of $CH_3OH$ [17,18]:

$$d = 10^4 \times \frac{Nm}{\rho \times 6.02 \times 10^{23}}$$

(Eq. 6)

Once deposited, a pre-irradiation FTIR absorbance spectrum of the ice was collected. The ice was then irradiated by 2 keV electrons for a total of 30 minutes, with spectra being collected at three-minute intervals. FTIR absorbance spectra were collected over the 4000-650 cm$^{-1}$ range at a resolution of 1 cm$^{-1}$ using 128 scans each measured over an integration time of 1 s. During every spectral acquisition, the electron beam was switched off.

During irradiation, electrons impacted the surface of the ice at an angle of 36°, as depicted in Fig. 1, and the beam spot (diameter = 8.4 mm) covered approximately 50% of the area monitored by the IR beam (diameter = 12 mm). The beam current (4.5 μA) was measured by making use of the collimator on the sample holder and the Faraday cup mounted opposite the gun and was used to calculate a flux of $4.21 \times 10^{14}$ electrons cm$^{-2}$ s$^{-1}$, as described in Sect. 3.1. Under such conditions, the power deposition of the beam is on the order of a few mW, and so it is expected that any physico-chemical changes within the ice as a result of heating are likely to be insignificant.

Electron irradiation of the $CH_3OH$ ice resulted in a decrease in each peak area of its characteristic absorbance bands, as well as the appearance of several new spectral features which were attributed to the formation of radiolytic products (Fig. 3; Table 1). These products included the molecules carbon monoxide (CO), carbon dioxide ($CO_2$), formaldehyde ($H_2CO$), and methane ($CH_4$). The formation of these products is in line with the results reported by prior studies of electron irradiation of frozen $CH_3OH$ [8,15,19-22].

Studies by Bennett *et al.* [21] and Schmidt *et al.* [22] have provided detailed information as to the likely formation routes of these molecules. Initial fragmentation of $CH_3OH$ may occur via the loss of hydrogen (yielding the neutral radicals $CH_3O$ and $CH_2OH$), the loss of oxygen (yielding $CH_4$), or the simultaneous (one-step) loss of two hydrogen atoms (yielding $H_2CO$). Abstraction of hydrogen from $CH_3O$ or $CH_2OH$ also affords $H_2CO$, which may itself undergo either successive (two-step) or simultaneous loss of two hydrogen atoms to yield CO. Reaction of CO with a hydroxide radical (OH) produces $CO_2$. Relevant rate constants for these processes were provided by Bennett *et al.* [21].

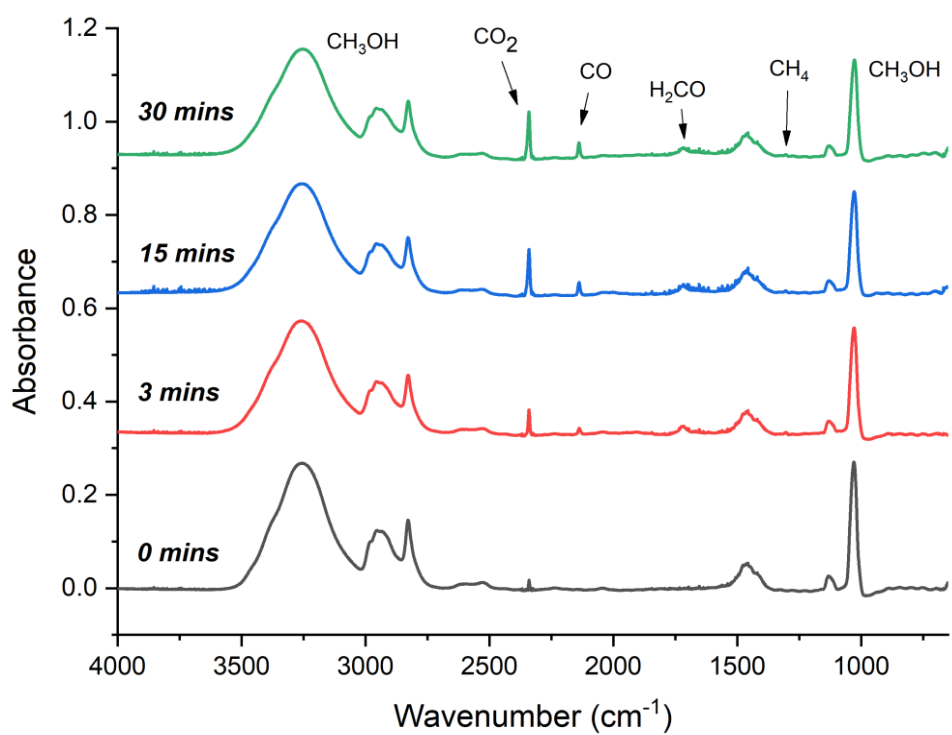

**Fig. 3** Spectral evolution of 1 μm $CH_3OH$ ice after irradiation by 2 keV electrons for 3, 15, and 30 minutes ($\Phi$ = $4.21 \times 10^{14}$ electrons cm$^{-2}$ s$^{-1}$). The formation of radiolytic product molecules has been highlighted. Spectra have been offset vertically for clarity.

**Table 1** List of FTIR absorbance bands used to identify the presence of molecular species in the $CH_3OH$ ice after 30 minutes irradiation with 2 keV electrons ($\Phi$ = $4.21 \times 10^{14}$ electrons cm$^{-2}$ s$^{-1}$).

| Peak Position [*integration limits*] (cm$^{-1}$) | Vibrational Mode Assignment | Integrated Band Strength ($10^{-17}$ cm molecule$^{-1}$) | Reference |
|---|---|---|---|
| 1027 [*989-1096*] | $v_8$ $CH_3OH$ | 1.610 | [18] |
| 1300 [*1294-1317*] | $v_4$ $CH_4$ | 0.776 | [23] |
| 1725 [*1689-1740*] | $v_4$ $H_2CO$ | 0.960 | [24] |
| 2138 [*2117-2162*] | $v_1$ CO | 1.100 | [25] |
| 2343 [*2315-2363*] | $v_3$ $CO_2$ | 7.600 | [26] |

We have analysed the column density evolution (as per Eqs. 5a and 5b outlined in Sect. 3.2) with increasing electron fluence for the $CH_3OH$ reactant, as well as for the product molecules CO, $CO_2$, and $CH_4$ (Fig. 4). Although the decay profile of the $CH_3OH$ column density does not reach an asymptotic limit at the end of irradiation, the irradiation period is nonetheless sufficiently long for Eqs. 5a and 5b to be considered a reasonable correction.

Our results show that the initial decay of amorphous $CH_3OH$ is fairly rapid, and only begins to slow down once a fluence of ~$2.50\times10^{16}$ electrons cm$^{-2}$ is reached. The production of CO is also most efficient at the earlier stages of irradiation, and begins to slow down at higher fluences due to its own radiolytic destruction or conversion to $CO_2$. The production of $CO_2$ follows a similar trend and its column density begins to plateau at a fluence of ~$2.50\times10^{16}$ electrons cm$^{-2}$, most likely due to the establishment of an equilibrium between its formation and conversion to other molecules such as CO or carbon trioxide ($CO_3$). Finally, the column density evolution of $CH_4$ follows a different pattern, with its abundance within the ice peaking fairly early on and subsequently diminishing under more prolonged irradiation.

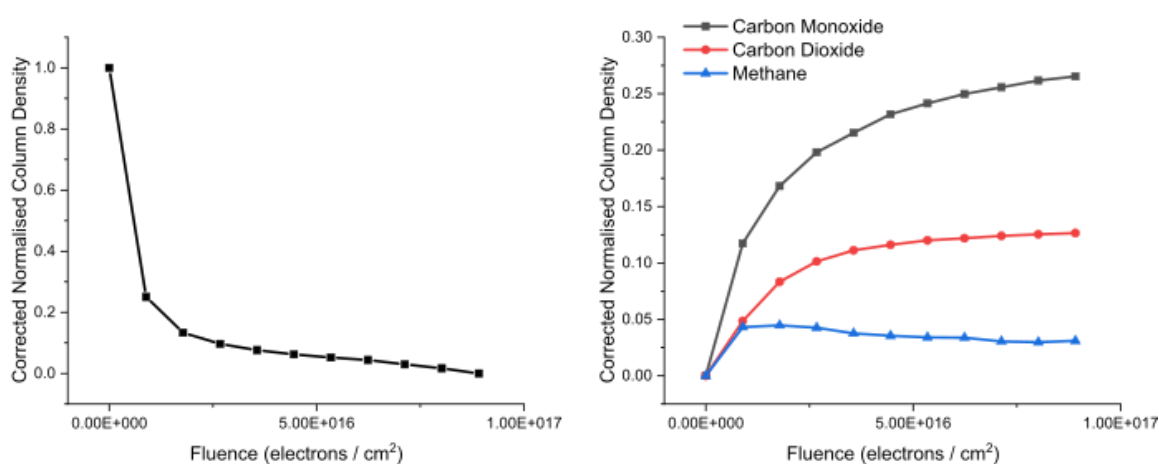

**Fig. 4** Column density evolution with increasing electron fluence for $CH_3OH$ (left) and product species (right) corrected using Eqs. 5a and 5b.

## 5 Thermal chemistry in a mixed ice containing $H_2O$ and $SO_2$

In this Sect., we outline the results of thermal reactions occurring in a mixed ice containing $H_2O$ and $SO_2$. Thermal reactions involving sulphur-bearing molecules (such as $SO_2$) are especially relevant to astrochemical and astrobiological studies of Solar System bodies such as Mars and the Galilean moons of Jupiter [27]. Previous studies have revealed that thermal reactions take place in mixed ices of $SO_2$ and $H_2O$, with the primary product being bisulphite ($HSO_3^-$) alongside a smaller quantity of meta-bisulphite ($S_2O_5^{2-}$) [28,29]. In the presence of oxidant species, the $SO_2$ may be oxidised to bisulphate ($HSO_4^-$) or sulphate ($SO_4^{2-}$) [30-33].

A 3 μm thick mixed $H_2O:SO_2$ ice (compositional ratio = 3:5) was deposited at 20 K onto a zinc selenide substrate, after which it was gradually warmed at a rate of 2 K min$^{-1}$ until a final temperature of 160 K was reached. FTIR absorbance spectra were collected at 20 K intervals. Results showed that, as the $H_2O:SO_2$ ice mixture was warmed, new peaks appeared centred at around 1040 cm$^{-1}$ and 956 cm$^{-1}$ (Fig. 5; Table 2). We ascribe the appearance of these peaks to the formation of $HSO_3^-$ and $S_2O_5^{2-}$, respectively [34,35]. The identification of these bands is only qualitative, as to the best of the authors' knowledge their integrated band strengths are not known, and so an accurate quantitative assessment of the amount of $HSO_3^-$ and $S_2O_5^{2-}$ produced cannot be performed. The chemical equations for the formation of these products are given below:

$$H_2O + SO_2 \rightarrow H^+ + HSO_3^-$$

(Eq. 7)

$$2\ HSO_3^- \rightarrow S_2O_5^{2-} + H_2O$$

(Eq. 8)

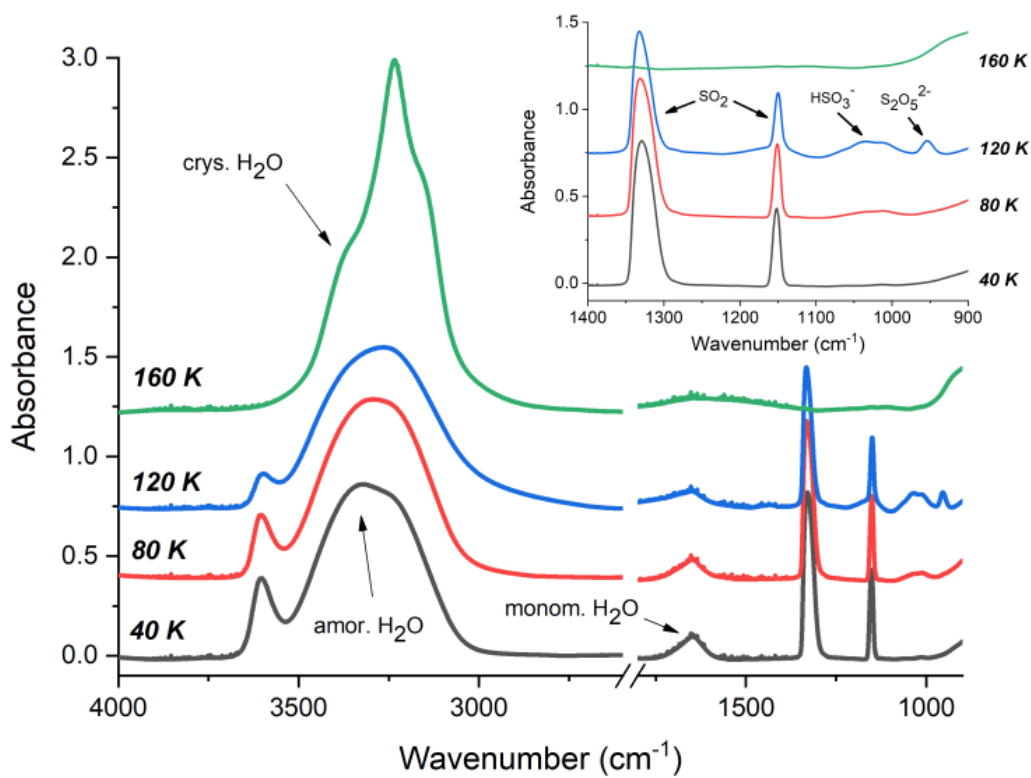

**Fig. 5** Spectral analysis of a thermally processed $H_2O$:$SO_2$ (compositional ratio = 3:5) mixed ice. As the ice is warmed, reactions between $SO_2$ and $H_2O$ result in the sequential formation of $HSO_3^-$ and $S_2O_5^{2-}$ (whose respective absorbance peaks at ~1040 cm$^{-1}$ and ~956 cm$^{-1}$ are more clearly visible in the inset). The formation of these products is most evident at 120 K, and further heating to 160 K causes a decline in the peaks attributable to the sulphur-containing species due to their sublimation, as well as a structural rearrangement of the water ice from the amorphous phase to the crystalline (hexagonal) one. In both the main figure and the inset, spectra are vertically offset for clarity. Increases in the 160 K spectrum at wavenumbers <1000 cm$^{-1}$ are due to changes in the high-temperature background profile that are not take into account by the background spectrum measured at 20 K.

The band area attributable to $HSO_3^-$ was initially visible at 40 K, and continued to grow in peak area with increasing temperature, representing a build-up in the concentration of this molecule within the ice structure (Figs. 5 and 6). The first diagnostic traces of $S_2O_5^{2-}$ were observed only later at 80 K. This is logical, as the formation of $S_2O_5^{2-}$ first requires sufficient accumulation of $HSO_3^-$ as per Eqs. 7 and 8. Further increases in temperature were accompanied by increases in the peak areas of both product molecules up until 120 K in the case of $HSO_3^-$ and 140 K in the case of $S_2O_5^{2-}$. At higher temperatures, peak areas for these molecules, as well as for $SO_2$, began to decrease due to sublimation [38]. By the time a temperature of 160 K was reached, there were no detectable traces of $SO_2$, $HSO_3^-$, or $S_2O_5^{2-}$ in the ice.

**Table 2** Vibrational mode assignments and characteristics for $SO_2$ and $H_2O$.

| Peak Position [*integration limits*] (cm$^{-1}$) | Vibrational Mode Assignment | Integrated Band Strength (10$^{-17}$ cm molecule$^{-1}$) | Reference |
|---|---|---|---|
| 1149 [*1136-1167*] | $v_1$ $SO_2$ | 0.22 | [36] |
| 1335 [*1274-1361*] | $v_3$ $SO_2$ | 1.47 | [36] |
| 1660 [*1586-1731*] | $v_2$ $H_2O$ | 1.20 | [26] |
| 3280 [*2964-3709*] | $v_1$, $v_3$ $H_2O$ | 14.00 | [37] |

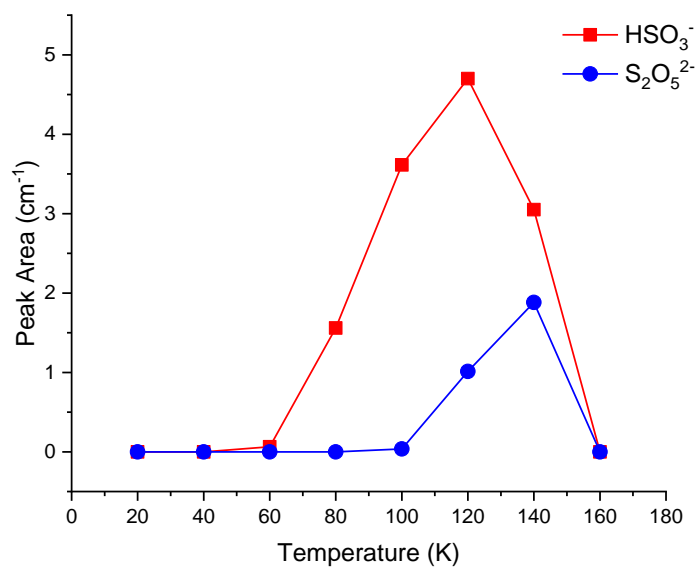

**Fig. 6** Peak area evolution of $HSO_3^-$ and $S_2O_5^{2-}$ with temperature. As the ice is warmed, reactions between $SO_2$ and $H_2O$ initially yield $HSO_3^-$, which then reacts to form $S_2O_5^{2-}$ as per Eqs. 7 and 8. Sublimation of all sulphur-containing molecules is achieved by 160 K. Conversion of the peak areas to molecular column densities as per Eq. 3 is not possible as the integrated band strengths $A_v$ for these molecules are not known.

## 6 Conclusions and future work

In this paper we have presented a technical and operational overview of the ICA facility located at the Institute for Nuclear Research (Atomki) in Debrecen with a particular emphasis on the characterisation of electron beams used for electron impact studies of astrophysical ice analogues. We have also validated results obtained by the facility through comparison of our preliminary experimental results with those obtained by other research groups.

The 2 keV electron irradiation of amorphous $CH_3OH$ ice at 20 K resulted in an initial rapid decay of $CH_3OH$ within the so-called active ice volume under consideration. Fragmentation of $CH_3OH$ led to the formation of new product molecules, including CO, $CO_2$, $H_2CO$, and $CH_4$, which could all be observed in the FTIR absorbance spectra. These results are in good agreement with the findings of several other studies considering the electron irradiation of $CH_3OH$, such as those by Bennett *et al.* [21] and Schmidt *et al.* [22].

Our thermal processing of a mixed $H_2O$:$SO_2$ ice of compositional ratio 3:5 demonstrated that, as the ice mixture was warmed from 20 K, molecular reactions resulted in the formation of new products, namely $HSO_3^-$ and $S_2O_5^{2-}$. The abundance of these products increased during warming until temperatures of 120 K ($HSO_3^-$) and 140 K ($S_2O_5^{2-}$) were reached, after which the sulphur-containing molecular components of the ice began to sublimate. These results mirror the findings of previous studies, particularly those by Moore *et al.* [28] and Kaňuchová *et al.* [29], who also considered such thermochemical reactions.

Further expansion of the capabilities of ICA is a core element of future planned work. At present, a viewport equipped with a laser interferometer for more accurate determination of the deposited ice layer thickness is under construction. An effusive evaporator will also be installed in the future to allow for the deposition and study of more refractory materials such as polycyclic aromatic hydrocarbons, biomolecules, and ionic compounds. We are also planning to increase the selection of spectrophotometers available for molecule identification, so as to bolster our current mid-IR capabilities with UV-vis measurements.

Finally, we note that at the time of writing ICA is a designated transnational access distributed planetary laboratory facility within the Europlanet Society consortium. As such, there are currently regular calls for research project proposals which are open to academic groups from around the world. Successful applications are funded by the European Union Horizon 2020 Research and Innovation Programme.

**Acknowledgements**

This article is based on work from the COST Action TUMIEE (CA17126) supported by COST (European Cooperation in Science and Technology). The authors all acknowledge funding from the Europlanet 2024 RI which has received funding from the European Union Horizon 2020 Research Innovation Programme under grant agreement No. 871149. The main components of ICA were purchased with the help of funding from the Royal Society through grants UF130409, RGF/EA/180306, and URF/R/191018. Support has also been received from the Hungarian Scientific Research Fund (grant No. K128621).

DVM is the grateful recipient of a University of Kent Vice-Chancellor's Research Scholarship. SI acknowledges the Royal Society for financial support. The research of ZK is supported by VEGA – the Slovak Grant Agency for Science (grant No. 2/0023/18) and the Slovak Research and Development Agency (contract No. APVV-19-0072). ATM thanks Queen Mary University of London for doctoral funding. The research of BP is supported by the European Union and the State of Hungary; co-financed by the European Regional Development Fund (grant GINOP-2.3.4-15-2016-00004).

**Author Contributions**

All authors contributed to the design, installation, and initial testing of ICA. DVM and BS wrote the manuscript, performed the experimental work, and performed the data analysis. ZJ, PH, and STSK performed the experimental